\documentclass[sigconf,anonymous=false]{acmart}

\AtBeginDocument{%
  \providecommand\BibTeX{{%
    \normalfont B\kern-0.5em{\scshape i\kern-0.25em b}\kern-0.8em\TeX}}}

\setcopyright{acmcopyright}
\copyrightyear{2022}
\acmYear{2022}
\acmDOI{XXXXXXX.XXXXXXX}

\usepackage{hyperref}
\usepackage{cleveref}
\usepackage{subcaption}
\usepackage{enumitem}
\usepackage{algorithm}
\usepackage{algpseudocode}
\begin{document}

\title[Research Track Paper]{AdaEnsemble: Learning Adaptively Sparse Structured Ensemble Network for Click-Through Rate Prediction}

\author{Yachen Yan}
\email{yachen.yan@creditkarma.com}
\orcid{1234-5678-9012}
\affiliation{%
    \institution{Credit Karma}
    \streetaddress{760 Market Street}
    \city{San Francisco}
    \state{California}
    \country{USA}
    \postcode{94012}
}

\author{Liubo Li}
\email{liubo.li@creditkarma.com}
\orcid{1234-5678-9012}
\affiliation{%
    \institution{Credit Karma}
    \streetaddress{760 Market Street}
    \city{San Francisco}
    \state{California}
    \country{USA}
    \postcode{94012}
}

\renewcommand{\shortauthors}{Yachen and Liubo}
\renewcommand{\subtitle}{AdaEnsemble}

\begin{abstract}
Learning feature interactions is crucial to success for large-scale CTR prediction in recommender systems and Ads ranking. Researchers and practitioners extensively proposed various neural network architectures for searching and modeling feature interactions. However, we observe that different datasets favor different neural network architectures and feature interaction types, suggesting that different feature interaction learning methods may have their own unique advantages. Inspired by this observation, we propose AdaEnsemble: a Sparsely-Gated Mixture-of-Experts (SparseMoE) architecture that can leverage the strengths of heterogeneous feature interaction experts and adaptively learns the routing to a sparse combination of experts for each example, allowing us to build a dynamic hierarchy of the feature interactions of different types and orders. To further improve the prediction accuracy and inference efficiency, we incorporate the dynamic early exiting mechanism for feature interaction depth selection. The AdaEnsemble can adaptively choose the feature interaction depth and find the corresponding SparseMoE stacking layer to exit and compute prediction from. Therefore, our proposed architecture inherits the advantages of the exponential combinations of sparsely gated experts within SparseMoE layers and further dynamically selects the optimal feature interaction depth without executing deeper layers. We implement the proposed AdaEnsemble and evaluate its performance on real-world datasets. Extensive experiment results demonstrate the efficiency and effectiveness of AdaEnsemble over state-of-the-art models.
\end{abstract}



\maketitle

\section{Introduction}
Click-through rate (CTR) prediction model~\cite{richardson2007predicting} is an essential component for the large-scale search ranking, online advertising and recommendation system~\cite{mcmahan2013ad,he2014practical,cheng2016wide,zhang2019deep}.

Many deep learning-based models have been proposed for CTR prediction problems in the industry. They have become dominant in learning the useful feature interactions of the mixed-type input in an end-to-end fashion\cite{zhang2019deep}. Although most of the existing methods can effectively capture higher-order feature interactions, we observe that their performance varies for different datasets. We believe this is due to their inductive bias: different methods learn different types of feature interactions and favor different datasets.

While every existing method focuses on automatically modeling different types of feature interactions, there have been few attempts to model different types of interactions jointly and dynamically. We believe that ensembling different interaction modules to create heterogeneous feature interactions can complement the non-overlapping knowledge that each interaction learning approach learned, as opposed to the homogeneous interaction modeling method, which restricts the types of feature interactions to be learned. For utilizing various interaction modules to learn different types of feature interactions, we use Sparsely-Gated Mixture-of-Experts (SpasrseMoE) architecture to enrich the model capacity while achieving computational efficiency through conditional computation.

We propose AdaEnsemble: a Sparsely-Gated Mixture-of-Experts (SparseMoE) hierarchical architecture to ensemble different interaction learning modules and dynamically select optimal feature interaction depth. Within each SparseMoE layer of AdaEnsemble, there is a collection of interaction learning experts, and a trainable gating network determines a sparse combination of these experts to use for each example. Within the Depth Selecting Controller, a trainable gating network will choose the feature interaction depth for each example and recursively propagate feature interaction representations through SparseMoE layers to the corresponding depth for computing the prediction. Through these conditional computation mechanisms, we enlarged the model capacity exponentially without increasing inference cost. The main contributions of this paper can be summarized as follows:
\begin{itemize}[leftmargin=10pt]
    \item We designed a novel model architecture called AdaEnsemble to ensemble various types of feature interaction learning modules by Sparsely-Gated Mixture-of-Experts (SparsseMoE). Through utilizing MoE layers recursively with residual connections and normalization, AdaEnsemble can model different types of interactions jointly and dynamically.
    \item We designed an efficient and effective Depth Selecting Controller to adaptively choose the optimal feature interaction depth. Through utilizing this controller, AdaEnsemble can dynamically determine the layer for early exiting to improve prediction accuracy and inference efficiency.
    \item We designed a bi-level optimization algorithm for iteratively training the modeling network and gating network.
    \item We conduct extensive experiments on real-world datasets and study the learning patterns of AdaEnsemble.
\end{itemize}

\section{Related Work}

\subsection{Feature Interaction Modeling}
Learning the feature interactions is the key topic in CTR prediction problems and has been widely discussed in literature. Various hybrid network architectures~\cite{cheng2016wide,qu2016product,qu2018product,wang2017deep,wang2021dcn,guo2017deepfm,lian2018xdeepfm} utilize the feed-forward neural network with non-linear activation function as its core component, to learn implicit interactions. The complement of the implicit interaction modeling improves the performance of the network that only models the explicit interactions~\cite{beutel2018latent}. 

Another group of models focuses on exploring bit-wise/vector-wise feature interactions. Deep \& Cross Network (DCN)~\cite{wang2017deep} and its improved version DCN V2  ~\cite{wang2021dcn} explores the feature interactions at the bit-wise level explicitly in a recursive fashion. Deep Factorization Machine (DeepFM)~\cite{guo2017deepfm} utilizes factorization machine layer to model the pairwise vector-wise interactions. Product Neural Network (PNN)~\cite{qu2016product,qu2018product} introduces the inner product layer and the outer product layer to learn vector-wise interactions and bit-wise interactions, respectively. xDeepFM~\cite{lian2018xdeepfm} learns the explicit vector-wise interaction using Compressed Interaction Network (CIN), which has an RNN-like architecture and learns vector-wise interactions using Hadamard product. FiBiNET~\cite{huang2019fibinet} utilizes Squeeze-and-Excitation network to dynamically learn the importance of features and model the feature interactions via bilinear function. AutoInt~\cite{song2018autoint} leverages the Transformer~\cite{vaswani2017attention} architecture to learn different orders of feature combinations of input features. xDeepInt~\cite{yan2020xdeepint} introduces polynomial interaction layer to recursively learn higher-order vector-wise and bit-wise interactions jointly with controlled degree, dispensing with jointly-trained DNN and nonlinear activation functions.

\subsection{Sparse Mixture-of-Experts Network}
The Sparsely-Gated MoE model~\cite{bengio2013estimating,shazeer2017outrageously} combines multiple experts and a trainable gating network that selects a subset of experts for each example. This network architecture can be viewed as a dynamic sparsity structure that maintains all weights but introduces sparsity into the model through conditional computation. The SparseMoE is widely used in natural language processing research area. Most of the discussions focus on improving the routing mechanism for the experts. Switch Transformer~\cite{fedus2021switch} simplifies the top-1 routing algorithm. GShard~\cite{lepikhin2020gshard} uses group-level top-2 routing. Both are trained with load balancing losses and improve language models with reduced communication and computational costs. BASE Layers~\cite{lewis2021base} treated routing as a linear assignment problem and removed the need for load balancing auxiliary losses. M6-T~\cite{yang2021m6} splits experts into different groups and applies k top-1 routing procedures. Some literature explore the training of the Sparsely-Gated MoE. EvoMoE~\cite{nie2021dense} decouples the training of experts and the sparse gate by training all experts at first and then gradually and adaptively becomes sparser while routes to fewer experts for learning the sparse gate. ST-MoE~\cite{zoph2022designing} further studies the training instabilities and uncertain quality issue of the MoE model. X-MoE~\cite{chi2022representation} proposed a dimension reduction and L2 normalization to solve the representation collapse in the training of MoE model.

\subsection{Early-Exiting Network}
The idea of early-exiting for the neural network was firstly proposed by BranchyNet~\cite{teerapittayanon2016branchynet} for computer vision. This technique is also applied to NLP tasks, DeeBERT~\cite{xin2020deebert}, FastBERT~\cite{liu2020fastbert}, and PABEE~\cite{zhou2020bert} was later introduced for improving inference efficiency of Transformer-Based BERT models.

For the early-exiting mechanism, BranchyNet~\cite{teerapittayanon2016branchynet}, DeeBERT~\cite{xin2020deebert}, FastBERT~\cite{liu2020fastbert} and SDN~\cite{kaya2019shallow} use the entropy-based or confidence-based criteria. While using entropy-based or confidence-based criteria is straightforward and effective, it takes advantage of the fact that the model's output is a probability distribution in multi-class classification tasks. This technique generally cannot be applied to binary classification and regression tasks. On the other hand, BERxiT~\cite{xin2021berxit} and Epnet~\cite{dai2020epnet} use learned modules for early-exiting.

\section{Proposed Model: AdaEnsemble}

\begin{figure}
    \centering
    \includegraphics[width=0.50\textwidth, height=0.50\textwidth]{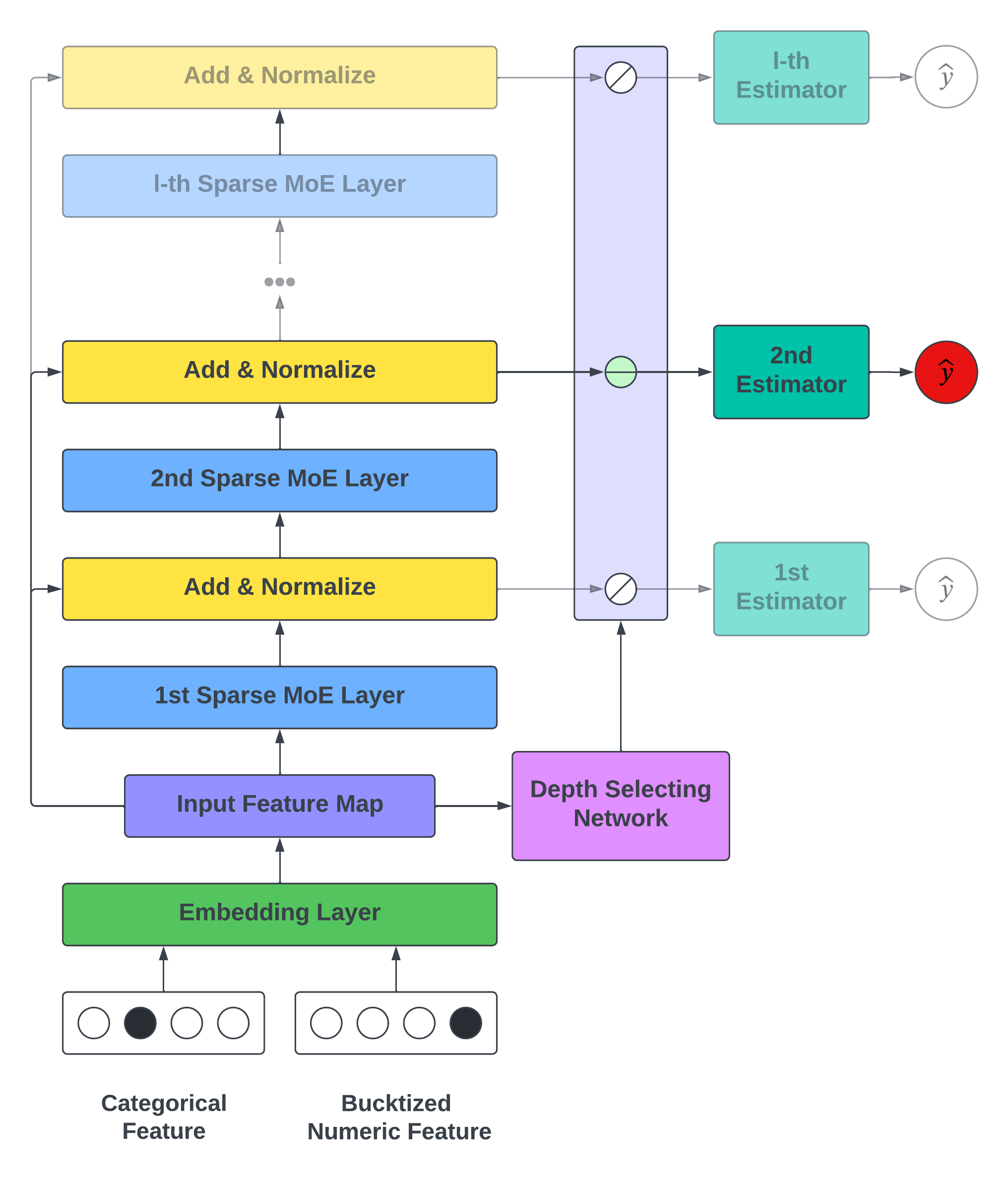}
    \caption{The Architecture of AdaEnsemble}
    \label{fig_adaensemble}
    \medskip
    \small
    In this example, the depth selecting network selects the 2nd layer to exit and compute the final prediction, therefore the deeper layers was not activated and plotted translucent in the figure.
\end{figure}

In this section, we give an overview of the architectures of AdaEnsemble. First, we introduce the feature processing and embedding layer, which maps continuous features and high-dimensional categorical features onto a dense embedding vector. Second, we introduce the feature interaction experts we considered for jointly learning the hierarchy of the deep feature representations. Third, we present the sparse mixture-of-experts (SparseMoE) layer, which ensemble multiple interaction experts dynamically, and the estimator associated with each SparseMoE Layer. Fourth, we discuss how to automatically and dynamically select the feature interaction depth based on the Depth Selecting Controller. Finally, a bi-level optimization algorithm will be provided for the training.

\subsection{Embedding Layer}

In large-scale CTR prediction tasks, inputs include both continuous and categorical features. Categorical features are often directly encoded by one-hot encoding, which results in an excessively high-dimensional and sparse feature space.

Suppose we have $F$ fields. In our feature processing step, we bucketize all the continuous features to equal frequency bins, then embed the bucketized continuous features and categorical features embed each feature onto a dense embedding vector $e_{i}$ of the same dimension $D$.

\begin{equation*}
    \mathbf{e}_{i} = \mathbf{x}_{i}\mathbf{V}_{i},
\end{equation*}
where ${e}_{i} \in R^{D}$, $\mathbf{V}_i$ is an embedding matrix for the $i$-th field, and $\mathbf{x}_{i}$ is the corresponding one-hot vector. Lastly, we concatenate $F$ embedding vectors and denote the output of embedding layer $X_0 \in R^{F\times D}$ as the input feature map:

\begin{equation}
\begin{aligned}
    X_0 = [e_{1}, e_{2}, \cdots, e_{F}]^\intercal.
\end{aligned}
\end{equation}

\subsection{Feature Interaction Experts}
We considered several types of feature interaction experts in our model: Dense Layer, Convolution Layer, Multi-Head Self-Attention Layer, Polynomial Interaction Layer, and Cross Layer. Essentially, any feature interaction learning layer can be included in our framework, and the residual connection and normalization will be applied to their ensembles. Now we introduce these feature interaction experts included in our framework. Note that our proposed framework is general and can use arbitrary feature interaction modules. The potential feature interaction experts can be used are not limited to the following.

\subsubsection{Dense Layer}
Dense Layer is also known as fully connected layer and is the most widely used module for modeling implicit feature interactions. In this paper, we use the dense layer with non-linear activation function for learning the deep feature representations. Given an input of embedding $X_{l-1}$, the output of embedding $X_{l}$ is obtained from:

\begin{equation}
\begin{aligned}
    X_{l} = \sigma(W_l \cdot X_{l-1})
\end{aligned}
\end{equation}

where $\sigma$ denotes activation function and $W_l$ denotes the weights of the $l$-th dense layer.

\subsubsection{Convolution Layer}
Convolution layers are widely used for computer vision problems. In this paper, we applied 1D convolution as one of the interaction experts. Here we utilize a dense layer ahead of the convolution layer for fusing the inputs embeddings first, as the convolution layer is locally connected. Given the embedding $X_{l-1}$ as input, the output of embedding $X_{l}$ is obtained from:

\begin{equation}
\begin{aligned}
    X_{l} = \text{Dense}(\text{Pooling}(\text{Conv1D}(\text{Reshape}(X_{l-1}))))
\end{aligned}
\end{equation}

Here we first reshape the input embedding and then apply 1D convolution followed by a pooling layer. Finally, we use a dense layer to project the output to the desired dimension.

\subsubsection{Multi-Head Self-Attention Layer}
Multi-Head Self-Attention Layer~\cite{vaswani2017attention} is widely used in transformer networks for its superior performance in natural language processing and has started to be popular in the computer vision research area. We consider utilizing Multi-Head Self-Attention Layer for modeling the dependency between features and forming meaningful higher-order features. Given an input of embedding $X_{l-1}$, the output of embedding $X_{l}$ is obtained from:

\begin{equation}
\begin{aligned}
    X_{l} = \text{Dense}(\text{MultiHeadSelfAttention}(\text{Reshape}(X_{l-1}))
\end{aligned}
\end{equation}

Here we first reshape the input embedding and then apply Multi-Head Self-Attention Layer followed by a dense layer to project the output to the desired dimension.

\subsubsection{Polynomial Interaction Layer}
Polynomial Interaction Network~\cite{yan2020xdeepint} is designed to capture bounded degree feature interactions explicitly. In this paper, we adopt the PIN layer as one of our feature interaction learning experts. Given an input of embedding $X_{l-1}$, the mathematical representation of the $l$-th PIN layer's output is given by:

\begin{equation}
\begin{aligned}
    X_{l} = X_{l-1} \circ (W_{l} \cdot X_{0})
\end{aligned}
\end{equation}

where $\circ$ denotes the Hadamard product and $W$ denotes the kernel weights of the PIN layer. We omit the residual connection from the original paper in the above equation as the residual connection will be used across the MoE layers.

\subsubsection{Cross Layer}
Deep Cross Network~\cite{wang2021dcn} is later proposed to explore the feature interactions in a recursive fashion. Given an input of embedding $X_{l-1}$, the output of embedding $X_{l}$ is obtained from:

\begin{equation}
\begin{aligned}
    X_{l} = X_{0} \circ (W_l \cdot X_{l-1}) + b_l
\end{aligned}
\end{equation}

Where $W$ and $b$ denote the weight matrix and bias vector in the $l$-th DCN layer. We also omit the residual connection of the original implementation in the above equation, as we will use the residual connection across the MoE layers.

\subsection{Sparse Mixture-of-Experts Layer}

\begin{figure}
    \centering
    \includegraphics[width=0.50\textwidth, height=0.40\textwidth]{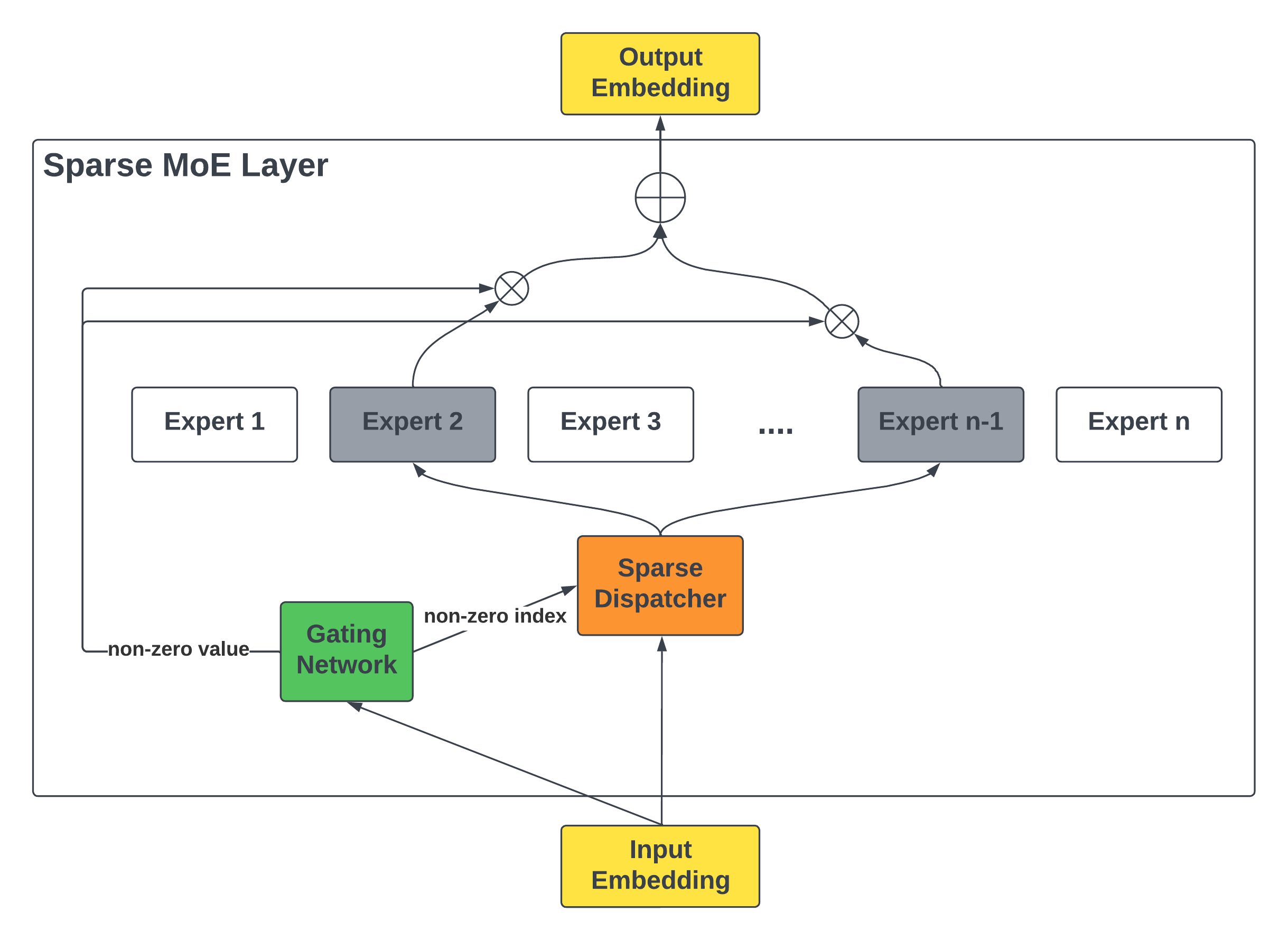}
    \caption{The architecture of Sparse Mixture-of-Experts Layer}
    \label{fig_sparsemoe}
    \medskip
    \small
\end{figure}

The Sparse Mixture-of-Experts layer ensembles aforementioned heterogeneous feature interaction experts and consists of several other essential parts to make the overall model can be stably trained.

\subsubsection{Noisy Gating Network}

\begin{figure}
    \centering
    \includegraphics[width=0.50\textwidth, height=0.45\textwidth]{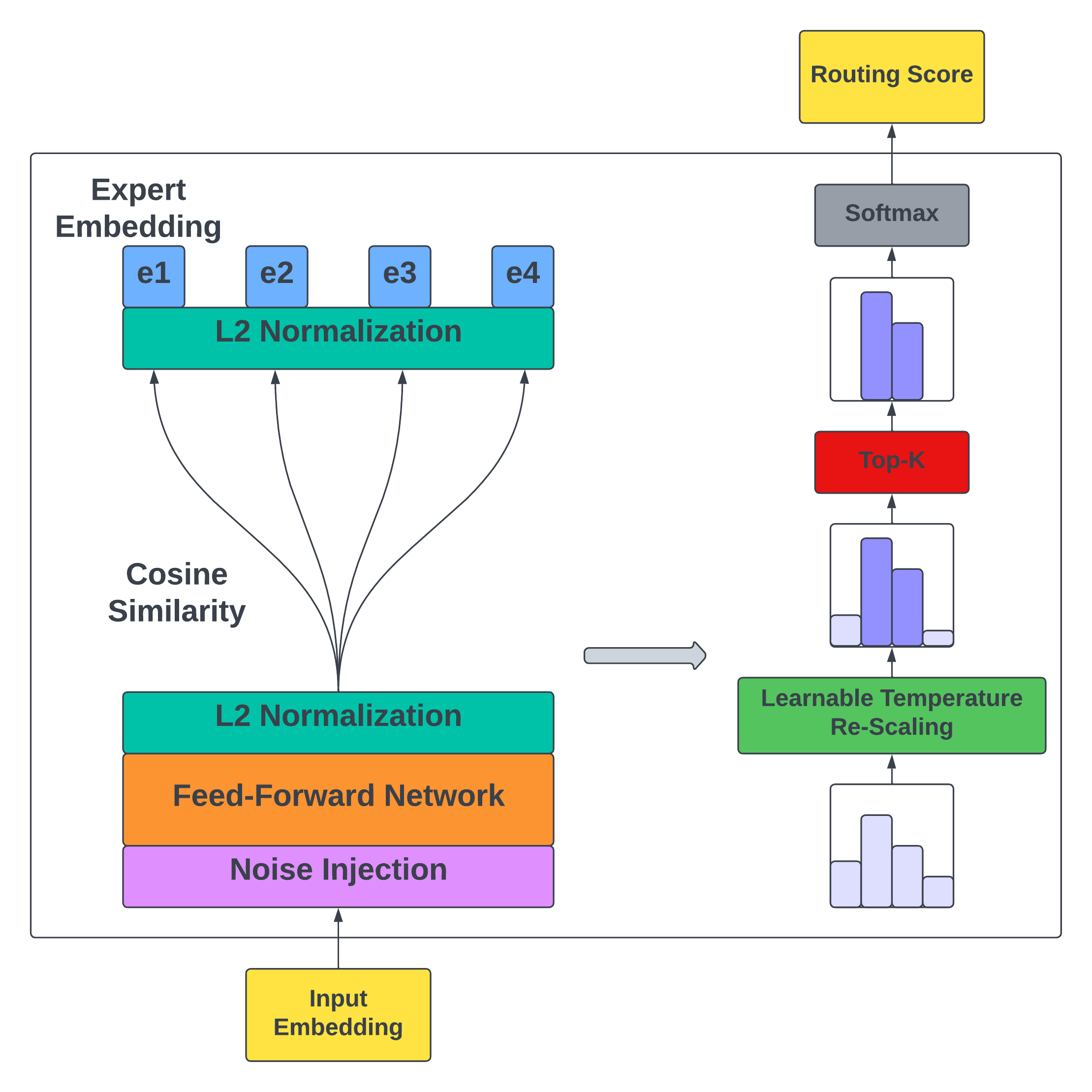}
    \caption{The Noisy Gating Network within Sparse Mixture-of-Experts Layer}
    \label{fig_gating_network}
    \medskip
    \small
\end{figure}

The gating network essentially computes the gating value for selecting and weighting the output embedding of each expert.

For the input embedding of gating network $X_{0}$, it firstly processed by the gating network: a two-layer feed-forward network, i.e. a dimension reduction layer with reduction ratio $r$~\cite{hu2018squeeze}, a non-linear activation function and then a dense layer projecting to hidden state $h \in {R}^{d}$. Additionally, we applied multiplicative jitter noise for introducing exploration and promoting load balancing between different experts.

\begin{equation}
\begin{aligned}
    h = \text{FFN}(X_0 \circ \text{RandomUniform}(1.0 - \text{eps}, 1.0 + \text{eps}))
\end{aligned}
\end{equation}

After projecting the input embedding to hidden state $h \in {R}^{d}$, we apply the $L_2$ normalization to both hidden state $h \in {R}^{d}$ and learnable expert embeddings $e_j \in {R}^{d}$, where $j$ is the index of expert. Then, we compute the cosine similarity between the hidden state and expert embedding as the initial routing score. Here we encourage the uniformity of representations to avoid dominated experts issue.

\begin{equation}
\begin{aligned}
    s_{j} = \frac{h \cdot e_{j}}{\Vert h \Vert \Vert e_{j} \Vert}
\end{aligned}
\end{equation}

Finally, we use a learnable temperature scalar $\tau$ to re-scale the routing scores to the range $[-1, +1]$.

\begin{equation}
\begin{aligned}
    g_{j} = s_{j} / \tau
\end{aligned}
\end{equation}

For the computed routing score $g$, we only keep the top k values and set the rest to $-\infty$, resulting in the corresponding softmax gating values equal $0$. The $i$-th element of the output of the gating network is
\begin{equation}
\begin{aligned}
    ExpertG(x)_i = \frac{\exp\Big( \text{TopK}(g, k)_i \Big)}{\sum_{j=1}^{N} \exp\Big( \text{TopK}(g, k)_{j} \Big)},
\end{aligned}
\end{equation}
where
\begin{equation}
\begin{aligned}
    \text{TopK}(g, k)_{j} =
    \begin{cases}
            g_{j} & \text{if $g_{j}$ is in the top $k$ elements of $g$} \\
            -\infty & \text{otherwise.}
        \end{cases}
\end{aligned}
\end{equation}

These gating values will be used by the sparse dispatcher for routing examples to different experts. This is the essential step for achieving sparsity of our Sparse Mixture-of-Experts layer. Note that the $G(x)$ is differentiable regardless the value of $k$\cite{fedus2021switch}.

\subsubsection{Annealing Top-K Gating}
We also introduce annealing mechanism to the Top-K operation. We starts with $k$ value equal to the number of experts, which means that we starts as a fully dense gate that routes examples to all experts. Then we gradually decrease the $k$ and route examples to fewer experts, to adaptively make the structure sparser and continuously improving the computation efficiency.

By annealing of the $k$ value, we start to train our architecture with a dense structure which allows us to thoroughly learn all experts and adjust the gating network in the correct direction at the beginning. Therefore, we can control the sparsity of our architecture while training to not only accelerate the convergence of the gating network but also benefit the experts' specialty for learning particular types of feature interactions.

\subsubsection{Sparse Dispatcher}
The sparse dispatcher takes the examples gating values and experts as input. It firstly dispatches the examples to the experts corresponding to the non-zero gating values, and lets experts generate the output embeddings. The output $y$ of the Sparse Mixture-of-Experts layer is the linearly weighted combination of expert output embeddings by the non-zero gating values.

\begin{equation}
\begin{aligned}
    y = \sum_{j \in \mathcal{\phi}} ExpertG_{j}(x) E_{j}(x)
\end{aligned}
\end{equation}

Where $\phi$ denotes the selected non-zero indices. We save computation based on the sparsity of $G(x)$. Wherever $G(x)_j=0$, we don't pass the expert to the corresponding expert and do not need to compute expert embedding $E_j(x)$.

\subsubsection{Load Distribution Regularization}
As stated in the previous research~\cite{shazeer2017outrageously,fedus2021switch,zoph2022designing,chi2022representation}, the gating network tends to select only a few experts if no regularization is applied, especially when certain experts are easier to train than other experts. This phenomenon is self-reinforcing, since the selected experts are trained more and will be selected more frequently by the gating network. Therefore, the load balancing loss is applied to enforce the uniform expert routing.

\begin{equation}
\begin{aligned}
    L_{\text{balance}} = \lambda \cdot N \cdot \sum_{j=1}^{N} f_{j} \cdot P_{j}
\end{aligned}
\end{equation}

where $N$ is the number of experts, $f_{j}$ is the fraction of examples dispatched to expert j, $P_{j}$ is the average of the router probability allocated for expert j, and $\lambda$ is the coefficient for the regularization term.

\begin{equation}
\begin{aligned}
    f_{j} = \frac{1}{B} \sum_{x \in \mathcal{B}} \mathbf{1} \{\text{argmax}\: p(x) = j\}
\end{aligned}
\end{equation}

\begin{equation}
\begin{aligned}
    P_{j} = \frac{1}{B}\sum_{x \in \mathcal{B}} p_{j}(x)
\end{aligned}
\end{equation}

While the default load balancing loss is applicable and effective when experts are of the same type, AdaEnsemble is using heterogeneous feature interaction experts, and the optimal load for each expert is not uniform. Therefore, we apply the below load distribution regularization to encourage the expected load distribution of heterogeneous experts.

\begin{equation}
\begin{aligned}
    L_{\text{distribution}} = \lambda \cdot \sum_{j=1}^{N} \frac{f_{j} \cdot P_{j}}{w_{j}}
\end{aligned}
\end{equation}

where $w_{j}$ is the expected load fraction of examples dispatched to expert j, and naturally $\sum_{j=1}^{N} w_{j} = 1$. In practice, the $\lambda$ should be sufficiently large to prevent expert selection self-reinforcing phenomenon at the initial training stage while not overwhelming the primary LogLoss objective.

\subsection{Estimator Layer}
The output of the Sparse Mixture-of-Experts layer is a feature map that consists of feature interactions of different degrees and types, including raw input feature map reserved by residual connections and higher-order feature interactions jointly learned by experts. For the final prediction, we merely use the formula as follows:
\begin{equation}
\begin{aligned}
    \hat{y} & = \sigma(W_{l}X_{l} + b_{l})
\end{aligned}
\end{equation}
where $\sigma$ is the sigmoid function, $W_{l}\in R^{1\times F}$ is a feature map aggregation vector that linearly combines all the learned feature interactions in the feature map, $b\in R$ is the bias.

\subsection{Depth Selecting Controller}

\subsubsection{Depth Selecting Network}
The Depth Selecting Network is essentially the same configuration as the aforementioned Noisy Gating Network for SparseMoE layer. We denote it by $DepthG(x)$. The outputs of $DepthG(x)$ are $[g^{depth}_{1}, g^{depth}_{2}, \cdots, g^{depth}_{L}]$, indicating each example's optimal forward propagation depth. The $l$-th unit denotes the probability of selecting the $l$-th MoE layer to exit. The optimal depth is automatically selected as the one corresponding to the largest probability. In contrast to the expert selection, when choosing the optimal depth of each example for the dynamic inference, we only keep the top-1 depth index from the output units of the Depth Selecting Network. Note that we can also apply the load distribution regularization to encourage the examples' propagation depth distribution.

\subsubsection{Dynamic Propagation Mechanism}
With the depth gates $g^{depth}_{l} \in [0,1]$ computed by Depth Selecting Network, we obtain the optimal depth for each example. If $g^{depth}_{l} = 0$, we recursively forward propagate examples through MoE layers and compute deeper representation until $g^{depth}_{l} = 1$ or reaching the final layer. If $g^{depth}_{l} = 1$,  the forward propagation will be stopped and the corresponding $l$-th estimator will compute the prediction. To efficiently process a batch of examples with different optimal propagation depths, we utilize algorithm \autoref{algo_dynamic_propagation} for dynamic forward propagation.

\floatname{algorithm}{Algorithm}
\begin{algorithm}
\small{
\caption{Dynamic Propagation}
\begin{algorithmic}[1]

\State $\texttt{DepthGates} \gets \texttt{DepthSelectingNetwork(x)}$
\State $\widehat{y} \gets \texttt{DynamicPropagation(x, {DepthGates}, depth=0)}$
\State \Return $\widehat{y}$

\State 
\Function{DynamicPropagation}{$\texttt{Inputs, Gates, Depth}$}
    \State $\texttt{Outputs = MoE(Inputs)}$
    \State $\texttt{Depth} \mathrel{+}= 1$
    \If{$\texttt{Depth}$ == Number of Layer}
        \State $\widehat{y} = \texttt{Estimator(Outputs)}$
    \Else
        \State $\texttt{g = Gates[:, Depth]}$
        \State $\texttt{Outputs\textsubscript{keep}, Outputs\textsubscript{exit} = Dispatch(Outputs, g)}$
        \State $\texttt{Gates\textsubscript{keep}, \_ = Dispatch(Gates, g)}$
        \State 
        \State $\widehat{y}\textsubscript{keep} = \texttt{DynamicPropagation(Outputs\textsubscript{keep}, Gates\textsubscript{keep}, Depth)}$
        \State $\widehat{y}\textsubscript{exit} = \texttt{Estimator(Outputs\textsubscript{exit})}$
        \State $\widehat{y} = \texttt{Combine(}\widehat{y}\textsubscript{keep}, \widehat{y}\textsubscript{exit}\texttt{)}$
    \EndIf
    \State \Return $\widehat{y}$
\EndFunction

\end{algorithmic}
\label{algo_dynamic_propagation}}
\end{algorithm}

\subsection{Training}

\subsubsection{Training Objective}
The loss function we use a linearly weighted combination of the Log Loss and the auxiliary load distribution regularization,

\begin{align}
     Loss & = L_{\text{LogLoss}} + \lambda_1 L^{\text{expert}}_{\text{distribution}} + \lambda_2 L^{\text{depth}}_{\text{distribution}}
\end{align}
where $\lambda_1$ and $\lambda_2$ are the coefficients for weighting the load distribution regularization.

\subsubsection{Bi-Level Optimization}
The optimization task for training the AdaEnsemble is to jointly optimize the parameters $W$, which stands for the expert layers and estimator layers, and $\alpha$, which represents the expert gating network and depth selecting network. Inspired by the DARTS~\cite{liu2018darts}, we apply bi-level optimization algorithm for training our model, where $\alpha$ is the upper-level parameters and $W$ is the lower-level parameters. We apply algorithm \autoref{algo_bi_level_opt} to optimize $W$ and $\alpha$ alternatively and iteratively.

\floatname{algorithm}{Algorithm}
\begin{algorithm}
\small{
\caption{Bi-Level Optimization for AdaEnsemble}
\raggedright
{\bf Input}: training examples with corresponding labels, step size $t$\\
{\bf Output}: well-learned parameters $\mathbf{W}^*$ and $\mathbf{\alpha}^*$
\begin{algorithmic}[1]

\While{not converged}
    \State Sample a mini-batch of validation data
    \State Updating $\mathbf{\alpha}$  by descending $\nabla_\mathbf{\alpha} \;\mathcal{L}_{val} \big(\mathbf{W} - \xi \nabla_{\mathbf{W}}\mathcal{L}_{train} (\mathbf{W}, \mathbf{\alpha}),\mathbf{\alpha}\big)$
    \State ($\xi = 0$ for first-order approximation)
    
    \For{$i\gets 1, t$}
        \State Sample a mini-batch of training data
        \State Update $\mathbf{W}$ by descending $\nabla_{\mathbf{W}}\mathcal{L}_{train} (\mathbf{W}, \mathbf{\alpha})$
    \EndFor
\EndWhile

\end{algorithmic}
\label{algo_bi_level_opt}}
\end{algorithm}

\subsection{Discussion on AdaEnsemble}
The combination of sparse experts routing at each SparseMoE layer and the depth selecting controller brings two merits to the proposed model. On one hand, the stacked sparseMoE layers allow the proposed model to leverage the exponential combinations of sparsely gated experts, which brings in more predicting power. On the other hand, the depth selecting controller enables the proposed model to learn the instance-ware model depth. It improves the efficiency during model serving. In the next section, we will illustrate the effectiveness of the proposed model through some experimental studies.

\section{Experiments}
In this section, we focus on evaluating the effectiveness of our proposed models and seeking answers to the following research questions::
\begin{itemize}[leftmargin=10pt]
	\item \textbf{Q1}: How does our proposed AdaEnsemble perform compared to each baseline in the CTR prediction problem?
	\item \textbf{Q2}: How does the SparseMoE layer perform compared to DenseMoE, which utilizes all feature interaction experts? Does the cascade of SparseMoE layers effectively capture different types of feature interactions?
	\item \textbf{Q3}: How does the depth selecting controller perform compared to a full-depth network? Does the early exiting mechanism achieve both effectiveness and efficiency?
	\item \textbf{Q4}: How do different hyper-parameter settings influence the performance of AdaEnsemble?
\end{itemize}

\subsection{Experiment Setup}

\subsubsection{Datasets}
We evaluate our proposed model on three public real-world datasets widely used for research.

\textbf{1. Criteo.}\footnote{https://www.kaggle.com/c/criteo-display-ad-challenge} Criteo dataset is from Kaggle competition in 2014. Criteo AI Lab officially released this dataset after, for academic use. This dataset contains 13 numerical features and 26 categorical features. We discretize all the numerical features to integers by transformation function $\lfloor Log\left(V^{2}\right) \rfloor$ and treat them as categorical features, which is conducted by the winning team of Criteo competition.

\textbf{2. Avazu.}\footnote{https://www.kaggle.com/c/avazu-ctr-prediction} Avazu dataset is from Kaggle competition in 2015. Avazu provided 10 days of click-through data. We use 21 features in total for modeling. All the features in this dataset are categorical features.

\textbf{3. iPinYou.}\footnote{http://contest.ipinyou.com/} iPinYou dataset is from iPinYou Global RTB(Real-Time Bidding) Bidding Algorithm Competition in 2013. We follow the data processing steps of~\cite{zhang2014real} and consider all 16 categorical features.

For all the datasets, we randomly split the examples into three parts: 70\% is for training, 10\% is for validation, and 20\% is for testing. We also remove each categorical features' infrequent levels appearing less than 20 times to reduce sparsity issue. Note that we want to compare the effectiveness and efficiency on learning higher-order feature interactions automatically, so we do not do any feature engineering but only feature transformation, e.g., numerical feature bucketing and categorical feature frequency thresholding.

\subsubsection{Evaluation Metrics}
We use AUC and LogLoss to evaluate the performance of the models.

\textbf{LogLoss} LogLoss is both our loss function and evaluation metric. It measures the average distance between predicted probability and true label of all the examples.

\textbf{AUC} Area Under the ROC Curve (AUC) measures the probability that a randomly chosen positive example ranked higher by the model than a randomly chosen negative example. AUC only considers the relative order between positive and negative examples. A higher AUC indicates better ranking performance.

\subsubsection{Competing Models}
We compare AdaEnsemble with following models: LR (Logistic Regression)~\cite{mcmahan2011follow,mcmahan2013ad}, FM (Factorization Machine)~\cite{rendle2010factorization}, DNN (Multilayer Perceptron), Wide \& Deep~\cite{cheng2016wide}, DeepCrossing~\cite{shan2016deep}, DCN (Deep \& Cross Network)~\cite{wang2017deep}, PNN (with both inner product layer and outer product layer)~\cite{qu2016product,qu2018product}, DeepFM~\cite{guo2017deepfm}, xDeepFM~\cite{lian2018xdeepfm}, AutoInt~\cite{song2018autoint}, FiBiNET~\cite{huang2019fibinet}, xDeepInt\cite{yan2020xdeepint} and DCN V2~\cite{wang2021dcn}. Some of the models are state-of-the-art models for CTR prediction problem and are widely used in the industry.

\subsubsection{Reproducibility}
We implement all the models using Tensorflow~\cite{abadi2016tensorflow}. The mini-batch size is 4096, and the embedding dimension is 16 for all the features. For optimization, we employ Adam~\cite{kingma2014adam} with learning rate is tuned from $10^{-4}$ to $10^{-3}$ for all the neural network models, and we apply FTRL~\cite{mcmahan2011follow,mcmahan2013ad} with learning rate tuned from $10^{-2}$ to $10^{-1}$ for both LR and FM. For regularization, we choose L2 regularization with $\lambda$ ranging from $10^{-4}$ to $10^{-3}$ for dense layer. Grid-search for each competing model's hyper-parameters is conducted on the validation dataset. The number of dense or interaction layers is from 1 to 4. The number of neurons ranges from 128 to 1024. All the models are trained with early stopping and are evaluated every 2000 training steps.

The setup is as follows for the hyper-parameters search of AdaEnsemble: The number of recursive feature interaction layers $l$ is searched from 1 to 4. For the number of selected experts $k$ per SparseMoE layer, the searched values are from 1 to 3. For the reduction ratio for both the expert gating network and depth selecting network, we search from 4 to 16. We use G-FTRL optimizer for embedding table and Adam for the model weights. For AdaEnsemble, as the performance will be generally better when using more experts or layers, we only report the one with fewer experts or layers used if its AUC difference is within $0.02\%$ compared to the ones using one more expert or layer.

\subsection{Model Performance Comparison (Q1)}

\begin{table}[H]
	\caption{Performance Comparison of Different Algorithms on Criteo, Avazu and iPinYou Dataset.}
    \label{tbl_model_performance}
	\centering
	\resizebox{1.0\linewidth}{!}{
		\begin{tabular}{ccccccc}
			\hline
			& \multicolumn{2}{c}{Criteo}        & \multicolumn{2}{c}{Avazu}         & \multicolumn{2}{c}{iPinYou}                      \\
			Model        & AUC             & LogLoss         & AUC             & LogLoss         & AUC             & LogLoss           \\
			\hline
			LR           & 0.7924          & 0.4577          & 0.7533          & 0.3952          & 0.7692          & 0.005605          \\
			FM           & 0.8030          & 0.4487          & 0.7652          & 0.3889          & 0.7737          & 0.005576          \\
			DNN          & 0.8051          & 0.4461          & 0.7627          & 0.3895          & 0.7732          & 0.005749          \\
			Wide\&Deep   & 0.8062          & 0.4451          & 0.7637          & 0.3889          & 0.7763          & 0.005589          \\
			DeepFM       & 0.8069          & 0.4445          & 0.7665          & 0.3879          & 0.7749          & 0.005609          \\
			DeepCrossing & 0.8068          & 0.4456          & 0.7628          & 0.3891          & 0.7706          & 0.005657          \\
			DCN          & 0.8056          & 0.4457          & 0.7661          & 0.3880          & 0.7758          & 0.005682          \\
			PNN          & 0.8083          & 0.4433          & 0.7663          & 0.3882          & 0.7783          & 0.005584          \\
			xDeepFM      & 0.8077          & 0.4439          & 0.7668          & 0.3878          & 0.7772          & 0.005664          \\
			AutoInt      & 0.8053          & 0.4462          & 0.7650          & 0.3883          & 0.7732          & 0.005758          \\
			FiBiNET      & 0.8082          & 0.4439          & 0.7652          & 0.3886          & 0.7756          & 0.005679          \\
			xDeepInt     & 0.8111          & 0.4408          & 0.7672          & 0.3876          & 0.7790          & 0.005567          \\
			DCN V2       & 0.8086          & 0.4433          & 0.7662          & 0.3882          & 0.7765          & 0.005593          \\
			\hline
			AdaEnsemble  & \textbf{0.8132} & \textbf{0.4394} & \textbf{0.7687} & \textbf{0.3865} & \textbf{0.7807} & \textbf{0.005550} \\
			\hline
	\end{tabular}}
\end{table}

The overall performance of different model architectures is listed in \Cref{tbl_model_performance}. We have the following observations in terms of model effectiveness:
\begin{itemize}[leftmargin=10pt]
	\item FM brings the most significant relative boost in performance while we increase model complexity compared to LR baseline. This reveals the importance of learning feature interactions.
	\item Models with more than two feature interaction modules generally perform better than models with only a single feature interaction module, indicating the importance of jointly learned feature interaction representation.
	\item The optimal feature interaction depth varies by feature interaction module type and when combined with different module types, indicating the necessity for dynamically combining different feature interactions on different interaction depths.
	\item AdaEnsemble achieves the best prediction performance among all models. Our model's superior performance could be attributed to the fact that AdaEnsemble jointly model various types of feature interactions by adaptively selecting the feature interaction experts combination and determining the optimal feature interaction depth by the controller.
\end{itemize}

\subsection{Feature Interaction Expert Selection Analysis (Q2)}
\begin{figure}
    \centering
    \includegraphics[width=0.50\textwidth, height=0.35\textwidth]{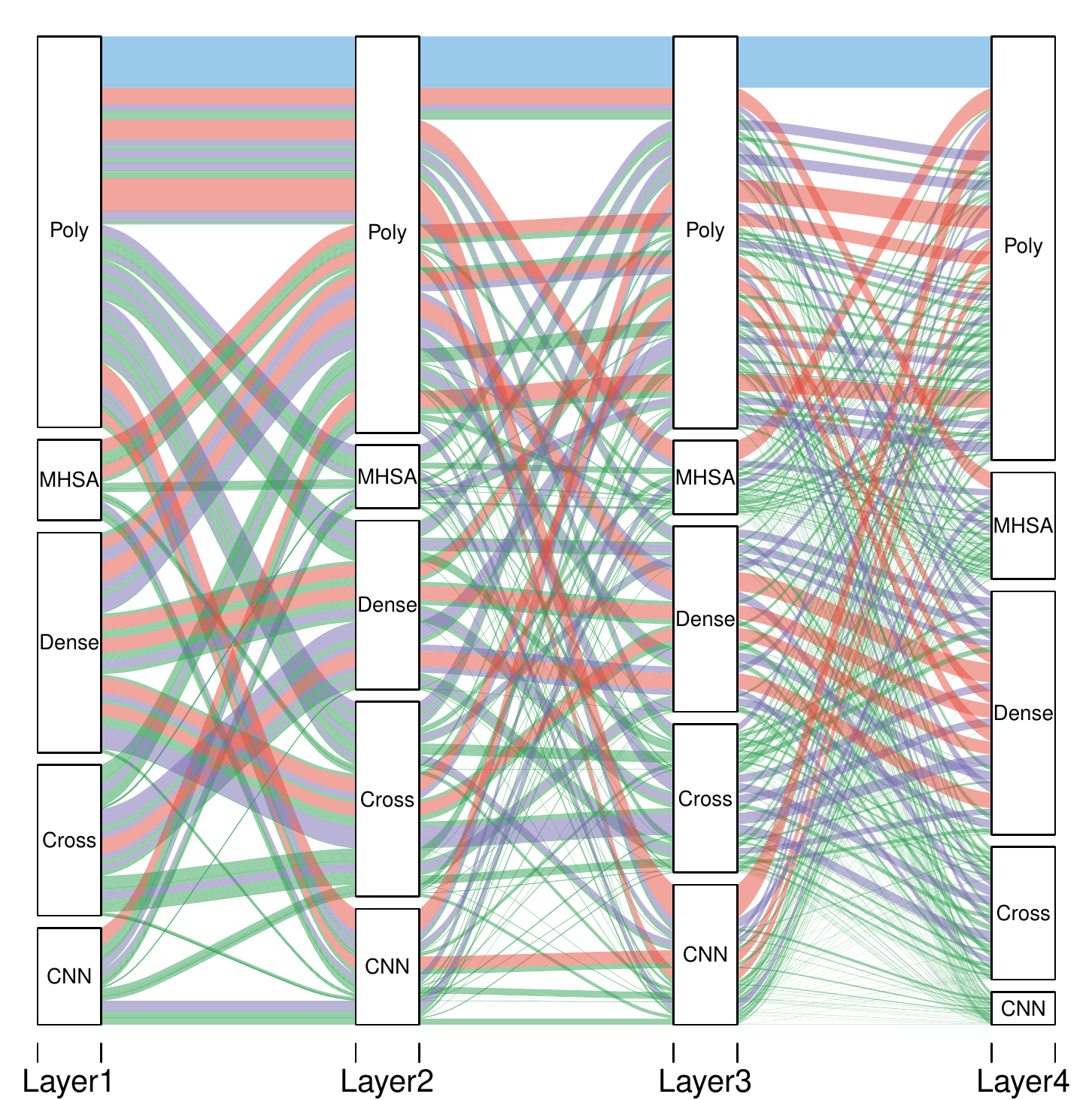}
    \caption{The Alluvial diagram for illustrating the dependency of each SparseMoE layer's expert selection}
    \label{fig_sparsemode_dependency}
    \medskip
    \small
    Each vertical axis represents a SparseMoE layer and the proportion of an expert being used. The horizontal flows indicate the dependency and relation of each SparseMoE layer's expert selection. The proportion of the expert combination was represented by the width of the flows and further clustered to different colors.
\end{figure}

We compare the model performance and FLOPs between the DenseMoE and SparseMoE layers in AdaEnsemble architecture. We also include the performance of different multi-layer single expert models and their ensemble. All the performance of above methods are listed in \Cref{tbl_expert_selection}. We also draw the alluvial diagram \autoref{fig_sparsemode_dependency} to illustrate the dependency of each SparseMoE layer's expert selection. The color of the flow is clustered by the frequency of the expert combination. Based on the above observations, we developed following understandings:

\begin{itemize}[leftmargin=10pt]
	\item Utilizing different feature interaction experts result in better performance than single expert models in general. SparseMoE layer achieves a better tradeoff between accuracy and computation efficiency.
	\item Only utilizing one expert per SparseMoE layer generally hurts the model performance as the model cannot ensemble different types of feature interactions.
	\item When utilizing more than one expert per SparseMoE layer, even though only a subset of feature interaction experts are selected, SparseMoE can still effectively capture the most significant feature interactions of different depths and maintain similar performance as the DenseMoE layer, while including more experts can also result in more computational cost.
	\item \autoref{fig_sparsemode_dependency} shows that the SparseMoE layers dynamically utilize a different combination of experts across different layers to capture the complex feature interactions effectively. That also explains why fusing different feature interactions is crucial for prediction accuracy.
\end{itemize}

\begin{table}[H]
	\caption{Performance Comparison of SparseMoE and DenseMoE on Criteo Dataset.}
    \label{tbl_expert_selection}
	\centering
	\begin{tabular}{c|ccc}
		\hline
                                    & AUC     & LogLoss & FLOPs     \\
		\hline
		SparseMoE(k=1)          & 0.8096  & 0.4423  & 2.26M     \\
		SparseMoE(k=2)          & 0.8121  & 0.4400  & 4.14M     \\
		SparseMoE(k=3)          & 0.8132  & 0.4394  & 6.02M     \\
		SparseMoE(k=4)          & 0.8133  & 0.4393  & 7.09M     \\
		DenseMoE                & 0.8133  & 0.4392  & 9.78M     \\
		Ensemble                & 0.8120  & 0.4401  & 12.15M    \\
		Dense Expert Only       & 0.8050  & 0.4463  & 3.71M     \\
		Cross Expert Only       & 0.8086  & 0.4433  & 3.36M     \\
		Polynomial Expert Only  & 0.8111  & 0.4408  & 3.32M     \\
		CNN Expert Only         & 0.8022  & 0.4501  & 1.11M     \\
		MHSA Expert Only        & 0.8051  & 0.4465  & 2.17M     \\
		\hline
	\end{tabular}
\end{table}

\subsection{Depth Selection Analysis (Q3)}
We compare the model performance between the AdaEnsemble with and without depth selecting controller to investigate whether the model achieves the harmony between prediction accuracy and inference efficiency with respect to depth selection. The performance of the different types of MoE layers and ensemble result is listed in \Cref{tbl_depth_selection}.

With the incorporation of the depth selecting controller, we can observe that our model can significantly improve training complexity and inference efficiency (measured in FLOPs) while achieving slightly better performance than the full-depth model. We think the full-depth model is easier to overfit compared to AdaEnsemble, thus resulting in slightly worse accuracy performance. The AdaEnsemble with depth selecting controller adaptively selects feature interaction depth per example basis, thus achieving better trade-offs between prediction accuracy and inference efficiency. The distribution of per example forward propagation depth is listed in \Cref{tbl_depth_distribution}.

\begin{table}[H]
	\caption{Performance Comparison of AdaEnsemble with and without controller on Criteo Dataset.}
    \label{tbl_depth_selection}
	\centering
	\begin{tabular}{c|ccc}
		\hline
                        & AUC     & LogLoss   & FLOPs     \\
		\hline
		w/ controller   & 0.8132  & 0.4394    & 6.02M     \\
		w/o controller  & 0.8128  & 0.4396    & 8.58M     \\
		\hline
	\end{tabular}
\end{table}

\begin{table}[H]
	\caption{AdaEnsemble Propagation Depth on Criteo Dataset.}
    \label{tbl_depth_distribution}
	\centering
	\begin{tabular}{c|cccc}
		\hline
                   & Layer 1   & Layer 2   & Layer 3   & Layer 4   \\
		\hline
		Fraction   & 6.53\%    & 19.36\%   & 66.43\%   & 7.68\%    \\
		\hline
	\end{tabular}
\end{table}

\subsection{Hyper-Parameter Study (4)}
In order to have deeper insights into the proposed model, we conduct experiments on the Criteo dataset and compare model performance on different hyper-parameter settings. This section evaluates the model performance change with respect to hyper-parameters that include: 1) depth of SparseMoE layers; 2) number of selected experts in SparseMoE layers;

\begin{figure}[htbp]
\centering
    \begin{subfigure}[b]{0.23\textwidth}  
    \includegraphics[width=\textwidth]{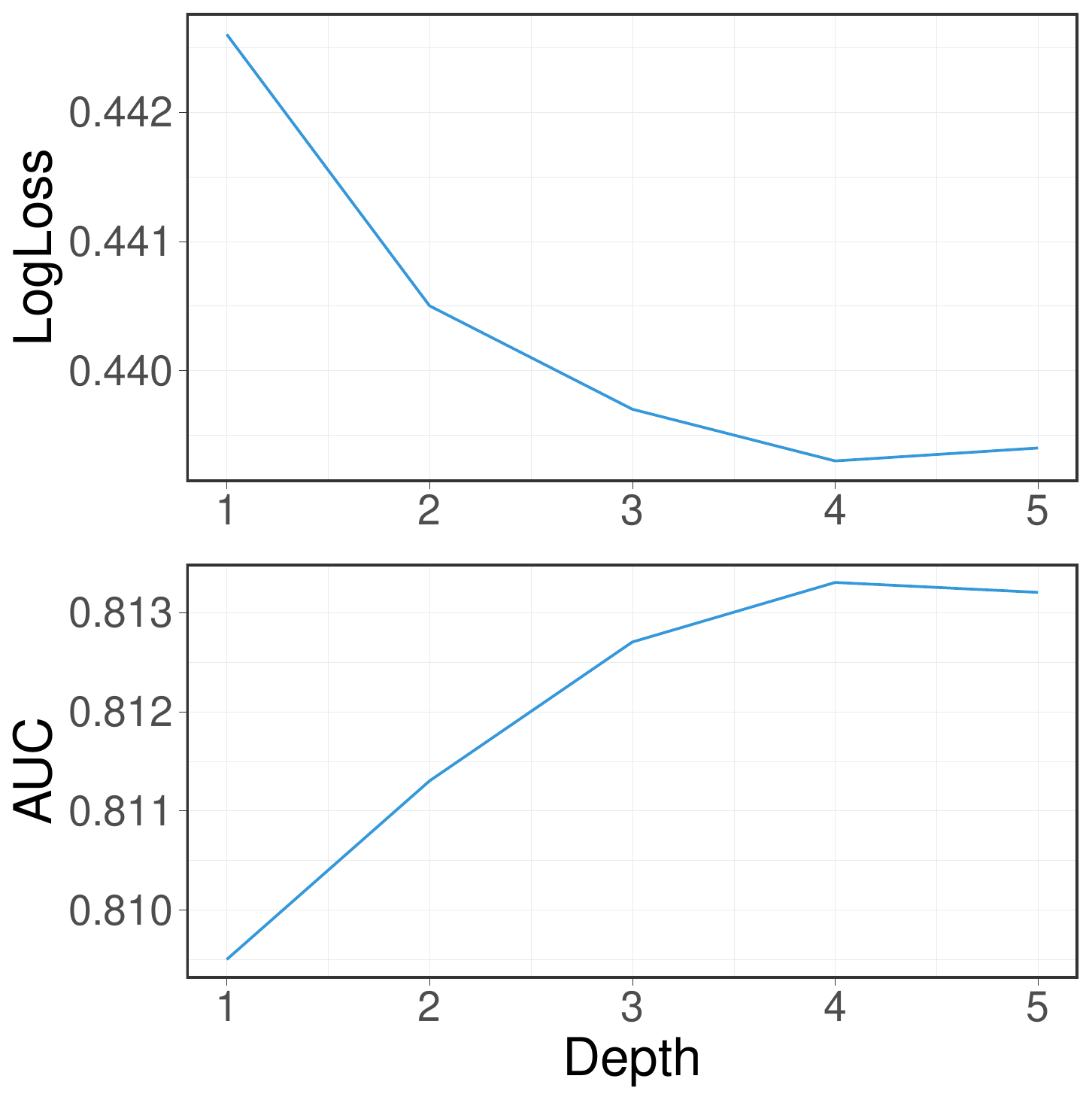}
    \caption{Layer Depth}
    \label{fig_depth}
    \end{subfigure}
    \hfill
    \begin{subfigure}[b]{0.23\textwidth}  
    \includegraphics[width=\textwidth]{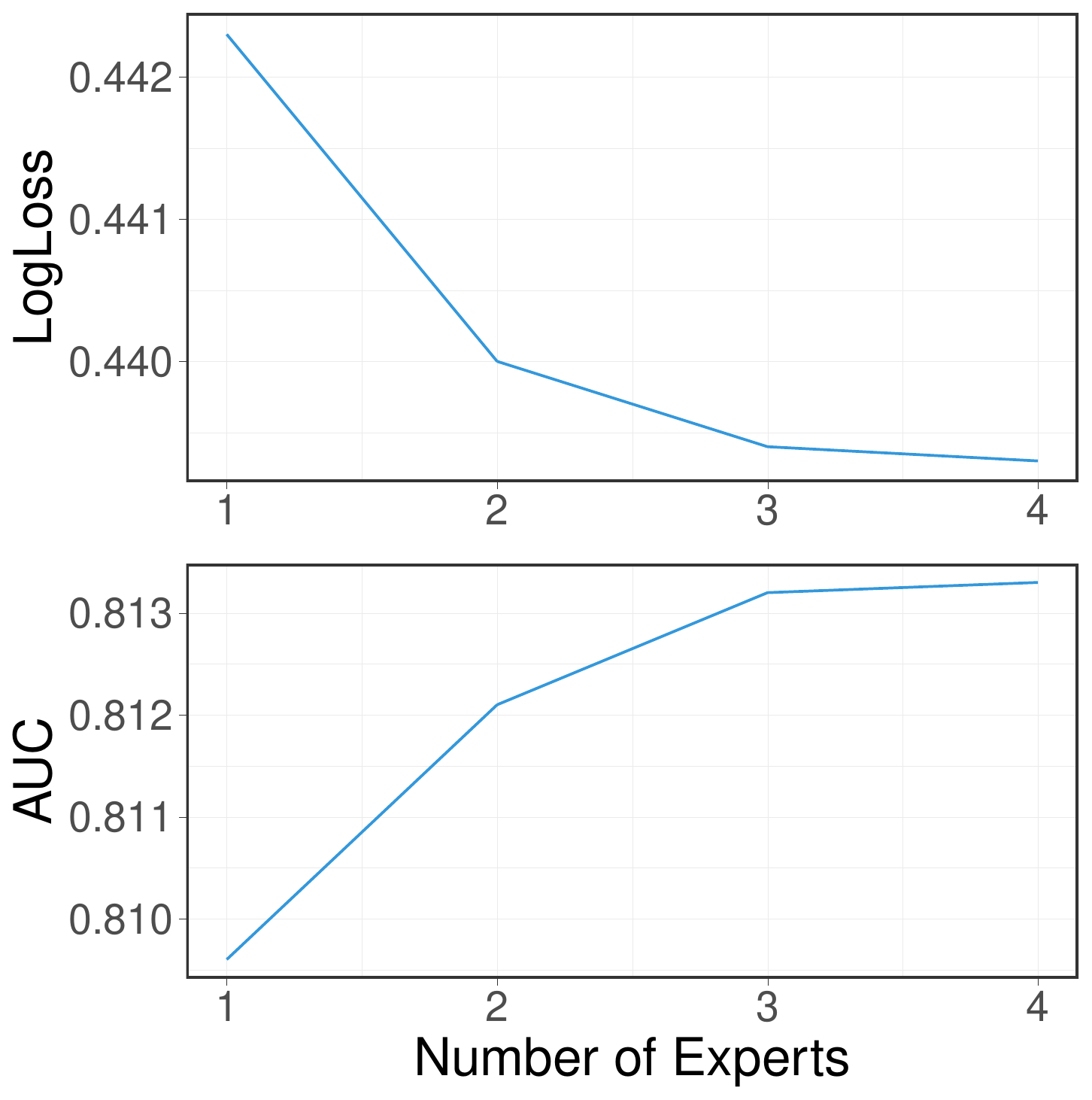}
    \caption{Number of Experts}
    \label{fig_num_experts}
    \end{subfigure}
\caption{Logloss and AUC v.s. feature interaction depth and number of experts.}
\label{fig_depth_num_experts}
\end{figure}

\subsubsection{Depth}
The depth of SparsMoE layers $l$ determines the maximum order of feature interactions learned. In this experiment, we set the number of selected experts $k$ as 3, which is generally a good choice for the Criteo dataset.

\autoref{fig_depth} shows the performance v.s. the depth $l$ of the AdaEnsemble on Criteo dataset. We observe that the performance keeps increasing until we increase the depth up to 4. This aligns with our understanding of the performance v.s. model complexity. Note that we still let the controller determine the interaction depth per example; the depth here is to control the maximum depth and model complexity.

\subsubsection{Number of Experts}
The number of selected experts of SparsMoE layers $k$ determines the number of selected feature interactions experts per SparseMoE layer. In this experiment, we set the depth of AdaEnsemble $l$ as 4, which is best for the Criteo dataset.

\autoref{fig_num_experts} shows the performance v.s. the number of experts $k$ for AdaEnsemble on Criteo dataset. We observe that the performance keeps increasing until $k$ equals 3. This indicates that the incremental gain diminishes while we increase the number of experts selected in SparseMoE layers.

\section{Conclusion}
In this paper, we proposed a new CTR model which ensembles the different interaction learning experts using the Sparse-Gated Mixture-of-Experts (SparseMoE) hierarchical architecture. We also introduce the Depth Selecting Controller for selecting the optimal depth for each example. Based on these two conditional computation mechanisms, our model will select a subset of experts and an optimal depth for each example. It enlarged the model capacity exponentially without increasing inference cost. Our comprehensive experiments have demonstrated the effectiveness and efficiency of our method.

In further work, We would like to study how to effectively extend our approach to user behavior sequence. While learning the sparse ensemble of different models, we expect our approach can dynamically select the optimal expert for different behaviors in the user behavior sequence data.

\newpage
\bibliographystyle{ACM-Reference-Format}
\bibliography{adaensemble.bib}

\end{document}